\documentclass[a4, 11pt]{article}
\usepackage[T1]{fontenc}

\usepackage{amssymb}        
\usepackage{latexsym}
\usepackage{times}
\usepackage{array}

\usepackage{graphicx}

\author{}

\date{}

\title{Better Approximation Algorithms for Maximum Asymmetric Traveling Salesman and Shortest Superstring}

\author{Katarzyna Paluch\\
Institute of Computer Science,  University of Wroc{\l}aw \\
{\tt abraka@cs.uni.wroc.pl}}

\newcommand{\dowod}{\noindent{\bf Proof.~}}
\newcommand{\koniec}{\hfill $\Box$\\[.1ex]}
\newtheorem{fact}{Fact}

\newcommand{\Ko}{{\cal K}_8}
\newcommand{\Kr}{{\cal K}_4}
\newcommand{\Kt}{{\cal K}_3}

\newtheorem{lemma}{Lemma}
\newtheorem{theorem}{Theorem}
\newtheorem{corollary}{Corollary}
\newtheorem{definition}{Definition}
\newtheorem{observation}{Observation}

\begin{document}

\maketitle
\thispagestyle{empty}
\begin{abstract}
In the maximum asymmetric traveling salesman problem (Max ATSP)  we are given a complete directed graph with nonnegative weights on the edges and we wish to compute a traveling salesman tour of maximum weight. 
In this paper we give a fast  combinatorial $\frac 34$-approximation algorithm for Max ATSP. It is based on a novel use of {\it half-edges}, matchings
and a new method of edge coloring. (A {\it half-edge} of edge $(u,v)$ is informally speaking ``either a head or a tail of $(u,v)$''.) 
The current best approximation algorithms for Max ATSP, achieving the approximation guarantee of $\frac 23$, are due to Kaplan, Lewenstein, Shafrir and Sviridenko  and Elbassioni, Paluch, van Zuylen. Using a recent result by Mucha, which states that an $\alpha$-approximation algorithm for Max ATSP implies a $(2+\frac{11(1-\alpha)}{9-2\alpha})$-approximation algorithm for the shortest superstring problem (SSP), we obtain also a $(2 \frac{11}{30} \approx 2,3667)$-approximation algorithm for SSP, beating the previously best known (having approximation factor equal to $2 \frac{11}{23} \approx 2,4782$.)
\end{abstract}

\newpage

\section{Introduction}
In the maximum asymmetric traveling salesman problem (Max ATSP)  we are given a complete directed graph $G=(V,E)$ with nonnegative weights on the edges and we wish to compute a traveling salesman tour of maximum weight. 
The problem is known to be APX-hard \cite{PY} and the current best approximation algorithms for it are due to Kaplan, Lewenstein, Shafrir and Sviridenko \cite{KLSS} and Elbassioni, Paluch, van Zuylen \cite{PEZ}. 
Both of them achieve the approximation ratio of $\frac 23$, the former is based on linear programming and the other is  combinatorial and simpler. 
Besides being an interesting problem in itself, Max ATSP is also of particular interest because of its applications to a number of related problems. For example,  an $\alpha$-approximation algorithm for Max ATSP  implies   a $(2+\frac{11(1-\alpha)}{9-2\alpha})$-approximation algorithm for SSP, which was recently shown by Mucha \cite{Mucha}. The shortest superstring problem is defined as follows. We are given $n$ strings $s_1, s_2, \ldots, s_n$ over a given alphabet $\sum$ and we want to find a shortest string $s$ such that each $s_i$ for $i, 1 \leq i \leq n$ is a  substring of $s$.  SSP arises in DNA sequencing and data compression. 
Currently the best approximation algorithm for SSP is due to Mucha \cite{Mucha} and achieves an approximation factor of $2 \frac{11}{23}$.
Any $\alpha$-approximation algorithm for Max ATSP implies also an algorithm with the same guarantee for the maximal compression problem defined by Tarhio and Ukkonen \cite{TU}.

 We devise a  combinatorial $\frac 34$-approximation algorithm for Max ATSP, thus proving
\begin{theorem}
There exists a $\frac{3}{4}$-approximation algorithm for the maximum tarveling salesman problem.
\end{theorem}
Using the result of Mucha \cite{Mucha}, we obtain
\begin{corollary}
 There exists a $2 \frac{11}{30}$-approximation algorithm for the shortest superstring problem.
\end{corollary}
The approach we have adopted is as follows. We start by   computing a maximum weight {\it cycle cover} $C_{max}$ of $G$, where
a cycle cover $C$ of graph $G$ is defined as a set of directed cycles of $G$ such that each vertex of $G$ belongs to exactly one cycle of $C$. A maximum weight cycle cover of $G$ can be found in polynomial time by a reduction to maximum weight matching.
Let $OPT$ denote the weight of a traveling salesman tour of $G$ of maximum weight.
The weight of an edge $e$ will be denoted as $w(e)$ and for any subset $E'$ of edges $E$ by $w(E')$ we will mean $\sum_{e \in E'} w(e)$. 
Since a traveling salesman tour is a cycle cover of $G$ (consisting of just one cycle), we know that  $w(C_{max}) \geq OPT$. By removing the lightest edge from each cycle of $C_{max}$, we obtain  a collection of vertex-disjoint
paths, which can be arbitrarily patched to form a tour. Removing the lightest edge from cycle $c$ of length $k$ results in a path of weight at least $\frac{k-1}{k} w(c)$.  Since $C_{max}$ may contain cycles of lenth two ($2$-cycles),
in the worst case the obtained tour may have weight equal to $\frac 12 w(C_{max})$. If we could find a maximum weight cycle cover of $G$ without cycles of length two ($2$-cycles) or three ($3$-cycles), then we would achieve a $
\frac 34$- approximation, but, unfortunately finding a maximum weight cycle cover without $2$-cycles is APX-hard \cite{BM}.

Since $2$- and $3$-cycles in a maximum weight cycle cover are an obstacle to getting a $\frac 34$-approximation, we would like to somehow get rid of them. The way we are going to achieve this is as follows. If $C_{max}$ contains at least one $2$-cycle or at least one $3$-cycle (a triangle), we compute a 
a cycle cover of $G$ that does not contain any $2$-cycle that already belongs to $C_{max}$  but  may contain {\bf \em half-edges} - a half-edge of edge $(u,v)$ is informally speaking
``either a head or a tail of $(u,v)$''. Such a cycle cover $C_1$ is going to be called a
{\bf \em relaxed cycle cover $\mathbf \mathit C_{1}$   improving $\mathbf \mathit C_{max}$}. Also we will ensure that a computed $C_1$ has weight at least $OPT$. 
 A similar relaxed cycle cover and half-edges have already been introduced in \cite{PEZ}. Computing $C_1$ is done via  a reduction to a maximum weight perfect matching. In some cases $C_1$ will suffice to build a traveling salesman tour of weight at least $\frac 34 OPT$. To (try to) extract such a tour
 from $C_1$ and $C_{max}$ we build a multigraph $G_1$ consisting of one copy of $C_{max}$ and two copies of $C_1$. Each occurrence of an edge $e$ in $C_{max}$ contributes one copy of $e$ to $G_1$ and each occurrence of $e$ in $C_1$ contributes two copies of $e$ to $G_1$. If $C_1$ contains only one half-edge of a certain edge $e$, then $C_1$ contributes one copy of $e$ to $G_1$.  The number of copies of edge $e$ in $G_1$ may be equal to up to three.
 The total weight of edges of $G_1$ is at least $3 OPT$. We would like to divide edges of $G_1$ into four sets $Z_1, Z_2, \ldots, Z_4$
 in such a way that each $Z_i$ ($1 \leq i \leq 4$) is a collection of
vertex-disjoint paths. One of the sets $Z_1, \ldots, Z_4$ would then have to have weight at least $\frac 34 OPT$ and by patching it to a tour, we would obtain the desired  solution. Dividing edges of $G_1$ into four sets
can be viewed as coloring them with four colors. We can immediately see that we are not able to color (the edges of) $G_1$ in the way described above if $C_1$ contains any one of the following: 
$(i)$ a $2$-cycle whose one edge belongs to some cycle of $C_{max}$ , $(ii)$ a triangle that already belongs to $C_{max}$  or a triangle that is oppositely oriented to some triangle of $C_{max}$, $(iii)$ a triangle whose one edge is contained in some $2$-cycle of $C_{max}$ or $(iv)$ a cycle of length $4$ such that two of its edges belong to two $2$-cycles of $C_{max}$. In the paper we will show that if $C_1$ contains neither of the above, then we are able
to color $G_1$ as required.

Otherwise, if $C_1$ contains one of the cycles that makes it impossible to color $G_1$ into four colors, we compute another cycle cover $C_2$ that does not contain: $(i)$ any $2$-cycle that already belongs to $C_{max}$,
$(ii)$ any  $2$-cycle that belongs to $C_1$ and whose one edge belongs to some cycle of $C_{max}$ , $(iii)$ any triangle $t$ such that both $C_{max}$ and $C_1$ contain a triangle on the same vertices  as $t$, $(iv)$ a triangle that belongs to $C_1$ and whose one edge is contained in some $2$-cycle of $C_{max}$ or $(v)$ a cycle of length $4$ that belongs to $C_1$  and such that two of its edges belong to two $2$-cycles of $C_{max}$ but may contain half-edges.
We call such a cycle cover  a {\bf \em relaxed cycle cover  improving $\mathbf \mathit C_{max}$  and $C_1$.} We will compute $C_2$ so that its weight is at least $OPT$.
Computing $C_2$ is done in a similar spirit to computing $C_1$ - via a novel reduction to a maximum weight perfect matching.

Once we have cycle covers $C_{max}, C_1$ and $C_2$ described above, we  build a multigraph $G_2$ consisting of two copies of each of these cycle covers. The total weight of edges of $G_2$ is at least $6 OPT$.
From $G_2$ we will want to extract a tour of weight at least $\frac 34 OPT$. To this end we will divide edges of $G_2$ into eight sets $Z_1, Z_2, \ldots, Z_8$ in such a way that each $Z_i$ ($1 \leq i \leq 8$) is a collection of
vertex-disjoint paths. One of the sets $Z_1, \ldots, Z_8$ will then have to have weight at least $\frac 34 OPT$.

The  task  of dividing edges of $G_1$ or $G_2$ into appropriately four or eight sets can equivalently be phrased as follows: we want to color each edge of $G_1$  or $G_2$ with one of four or correspondingly one of eight colors in such a way that each color class  consists of vertex-disjoint paths.For coloring both $G_1$ and $G_2$ we present a new method, which we think is interesting in its own right. 
One of the surprisingly simple ideas on which this method is based is as follows: let $S$ be a subset of $V$ and $e=(u,v)$ an edge going into $S$ (i.e. $u \notin S$ and $v \in S$), which is colored with a color $k$. Then if there exists no edge $e'=(u',v')$ outgoing from $S$ (i.e. such that $u' \in S$ and $v' \notin S$) which is colored $k$, then $e$ does not belong to any  cycle, whose all edges are colored $k$.  Another ingredient of the method are alternating cycles (a notion from matching theory) with certain properties.

{\bf Previous and related results}
The history of approximating the problems of maximum asymmetric traveling salesman and shortest superstring is quite long as prove the following lists of papers  \cite{Li},  \cite{Blum},  \cite{Teng}, \cite{Czumaj},  \cite{KPS}, \cite{armen95}, \cite{armen95}, \cite{BJJ}, \cite{sweedyk}, \cite{KLSS}, \cite{PEZ}, \cite{Mucha} and \cite{FNW}, \cite{KPS} \cite{B1}, \cite{LS}, \cite{KLSS},
\cite{PEZ}.

Other variants of the maximum traveling salesman problem that have been considered are among others: the maximum symmetric traveling salesman problem (MAX TSP), in which the underlying graph is undirected - currently the best known approximation ratio is $\frac 79$ \cite{P}, the maximum metric symmetric traveling salesman problem, in which the edge weights satisfy the triangle inequality - the best approximation factor is $\frac 78$ \cite{KM2},
the maximum asymmetric traveling salesman problem with triangle inequality - the best approximation ratio is $\frac{35}{44}$ \cite{KM1}.
 
\section{New upper bounds - relaxed cycle covers  improving $C_{max}$}
Suppose we have computed a maximum weight cycle cover $C_{max}$ of a given complete directed graph $G=(V,E)$. 
We will say that a cycle $c \in C_{max}$ is {\bf \em hard} if each edge $e$ of $c$ satisfies $w(e)>\frac{1}{4} w(c)$. We are going to call cycles of length $i$, i.e. consisting of $i$ edges, {\bf \em $i$-cycles}. Also sometimes $3$-cycles will be called {\bf \em triangles}. 
Let us notice that only $2$-cycles and triangles can be hard.
If $C_{max}$ does not contain a hard cycle, then we can easily build a traveling salesman tour of weight at least $\frac{3}{4} w(C_{max}) \geq \frac{3}{4} OPT$. 
If $C_{max}$ contains at least one hard cycle, we would like to obtain two other cycle covers $C_1$ and $C_2$ which may contain {\bf \em half-edges} - a half-edge of edge $(u,v)$ is informally speaking
``either a head or a tail of $(u,v)$''.

Below we give the precise definitions of a relaxed cycle cover  $C_1$ improving $C_{max}$  and a relaxed cycle cover improving $C_{max}$ and $C_1$.

\begin{definition}\label{rel}
Given  a complete directed graph $G=(V,E)$ with edge weights $w(u,v)\ge 0$ for every $(u,v)\in E$  and a maximum weight cycle cover $C_{max}$of $G$, let $\tilde G=(\tilde V, \tilde E)$ be the graph obtained from $G$ by replacing each $(u,v)\in E$ by a vertex $x_{(i,j)}$ and two edges $(u,x_{(u,v)})$ and $(x_{(u,v)},v)$, each with weight $\frac 12 w(u,v)$. Edges $(u, x_{(u,v)}),  (x_{(u,v)},v)$ will be called
{\bf \em half-edges (of $(u,v)$)}.
A {\bf \em relaxed cycle cover   improving $C_{max}$}  is a subset $\tilde C\subseteq \tilde E$ such that
\begin{itemize}
\item[(i)]
each vertex in $V$ has exactly one outgoing and one incoming half-edge in $\tilde C$;
\item[(ii)]
for each $2$-cycle $c$ of $C_{max}$  on vertices $u,v$ $\tilde C$ contains at most two half-edges from 
\newline $\{(u, x_{(u,v)}), (x_{(u,v)}, v), (v, x_{(v,u)}), (x_{(v,u)}, u)\}$.  Moreover if $\tilde C$ contains  one half-edge of $(u,v)$ and one half-edge of $(v,u)$, then one of them is incident with $u$ and the other with $v$.
\end{itemize}

Let $C_1$ be any relaxed cycle cover improving $C_{max}$. 

A {\bf \em relaxed cycle cover  improving $C_{max}$ and $C_1$} is a subset $\tilde C\subseteq \tilde E$ such that 
\begin{itemize}
\item[(i)] each vertex in $V$ has exactly one outgoing and one incoming half-edge in $\tilde C$; 
\item[(ii)] for each $2$-cycle $c$ on vertices $u,v$ such that $c$ either belongs to $C_{max}$ or $c$ belongs to $C_1$ and one of its edges belongs to some cycle of $C_{max} \ $  $\tilde C$ contains at most two half-edges from 
\newline $\{(u, x_{(u,v)}), (x_{(u,v)}, v), (v, x_{(v,u)}), (x_{(v,u)}, u)\}$.  Moreover if $\tilde C$ contains  one half-edge of $(u,v)$ and one half-edge of $(v,u)$, then one of them is incident with $u$ and the other with $v$; 
\item[(iii)] for each  triangle $t$  on vertices $p,q,r$ containing edges $(p,q), (q,r), (r,p)$ such that either (1) $t \in C_{max} \cap C_1$ or (2) $t \in C_{max}$ and $C_1$ contains a $2$-cycle on two vertices of $t$ or
(3) $t \in C_1$ and $C_{max}$ contains a $2$-cycle on two vertices of $t$,

  $\tilde C$ contains at most four and the even number of half-edges  of edges of $t \cup t'$, where $t'$ denotes a triangle oppositely oriented to $t$.  Also for any edge $(v_1, v_2)$, where $(v_1,v_2) \in \{(q,p), (r,q), (p,r)\}$,  $\tilde C$ satisfies the following:
if $w(v_1,v_2) > \frac 12 (w(t)-w(v_2,v_1))$, then  $\tilde C$ does not contain all half-edges from the set $\{(v_1, x_{(v_1,v_2)}),  (x_{(v_1,v_2)}, v_2), (v_3, x_{(v_3,v_2)}), (x_{(v_2,v_3)}, v_3)\}$, where $v_3$ belongs to  $\{p,q,r\} \setminus \{v_1, v_2\}$; 
\item[(iv)]  let $c$ be a cycle of length $4$ containing edges $(p,q), (q,r), (r,s), (s,p)$  and such that $c$ either belongs to $C_{max}$ and $C_1$ has two $2$-cycles that share an edge with $c$ or vice versa. Then for each such $c$ $\tilde C$
contains at most six and the even number of half-edges of edges of $c \cup c',$ where $c'$ denotes a $4$-cycle oppositely oriented to $c$.
\end{itemize}
\end{definition}

\begin{figure}[ht]
\begin{minipage}[b]{0.45\linewidth}
\centering
\includegraphics[scale=0.5]{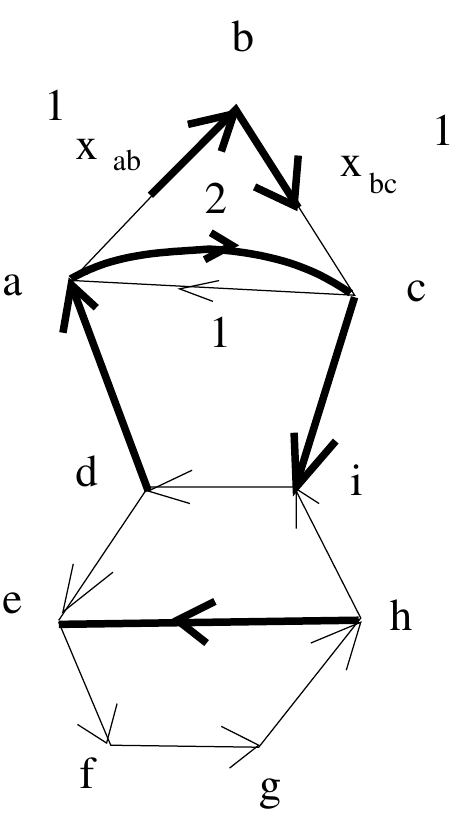}
\caption{{\scriptsize 
$C_{max}$ consists of triangle $abc$ and $6$-cycle $defghi$. $\tilde{C}$ consists of two $4$-cycles $acid$, $efgh$ and two half-edges $(x_{(ab)},b)$ and $(b, x_{(bc))}$. The only edges with positive weight are: $(ab), (bc), (ca), (a,c)$. Thus $w(C_{max})= w(\tilde{C})=3$. Also edge $(a,c)$ is such that $w((a,c)) > \frac 12 (3-1)$.  A maximum traveling salesman tour in this graph has weight $2$, which means that $\tilde{C}$ is useless in getting an approximation better than $\frac 23$. 
}}
\label{creldef}
\end{minipage}
\hspace{0.5cm}
\begin{minipage}[b]{0.45\linewidth}
\centering
\includegraphics[scale=0.6]{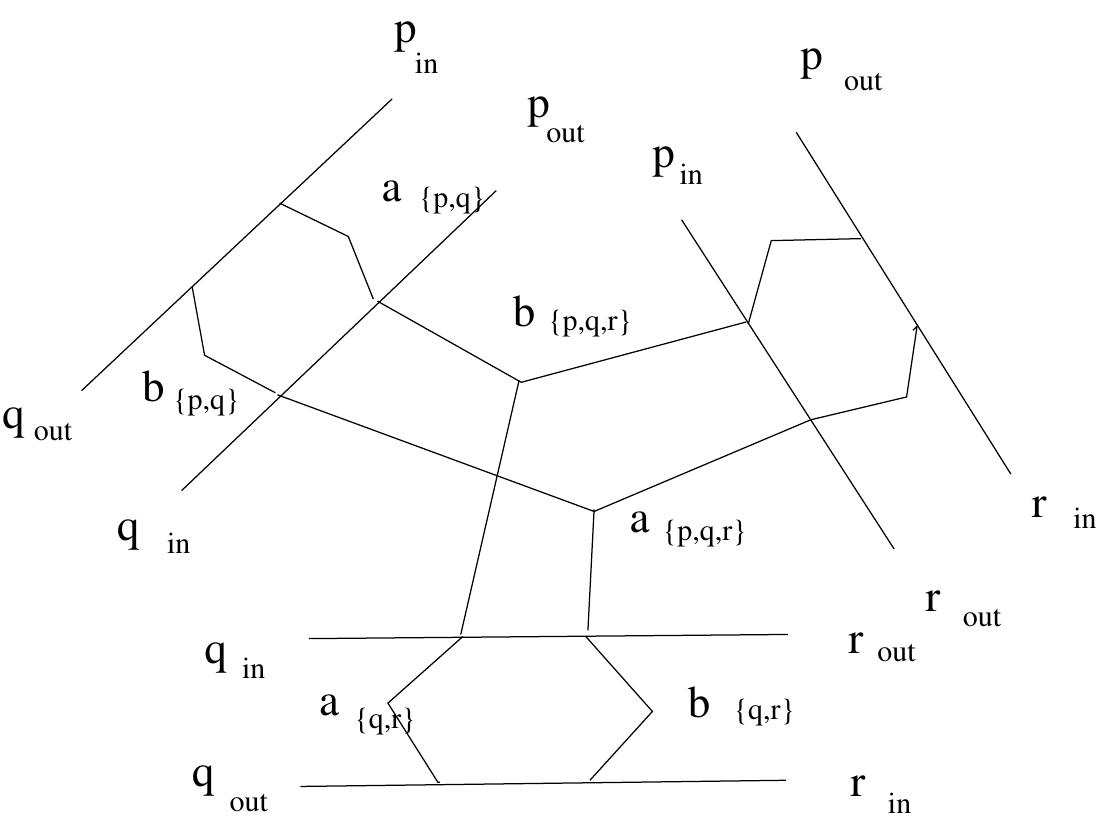}
\caption{{\scriptsize A gadget corresponding to a triangle $t$ on vertices $p,q,r$ such that $t$ consists of edges $(p,q), (q,r), (r,p)$ and $w((p,r)) \leq \frac{1}{2}(w((p,q))+w((q,r))$. To make the figure more readable
vertices $p_{in}, p_{out}, q_{in}, q_{out}, r_{in}, r_{out}$ are drawn twice each. 
Vertices $a_{\{p,q,r\}}, b_{\{p,q,r\}}$ are connected correspondingly to vertices $e^2_{pq}, e^1_{rq}, e^1_{rp}$ and $e^1_{pq}, e^2_{rq}, e^2_{rp}$. Notice that  $a_{\{p,q,r\}}, b_{\{p,q,r\}}$ are connected to vertices on edge
$(r,q)$ which does not belong to $t$.}}
\label{gtriangle}
\end{minipage}
\end{figure}

In graph $\tilde{G}$ each original edge of $G$ is replaced by two half-edges. Since this time on, by saying an edge, we will mean an edge of $G$ and by saying a half-edge we will mean an edge of $\tilde{G}$.
Also by  saying that an edge $(u,v)$ of $G$ belongs to $\tilde{C}$, we will mean that both half-edges of $(u,v)$ belong to $\tilde{C}$.
Condition $(ii)$ in the above definitions ensures that $\tilde{C}$ does not contain certain $2$-cycles.  The first part of condition $(iii)$ says that $\tilde{C}$ does not contain all the edges of a given triangle or a triangle oppositely oriented.  The reason for having the second part of  condition $(iii)$, in which we forbid the given subsets of half-edges,
is illustrated in Figure \ref{creldef}. 

We will say that a cycle $c$ of $C_1$ is {\bf \em problematic} if it is a $2$-cycle that shares an edge with some cycle of $C_{max}$ or it is a triangle or a $4$-cycle  having properties described in conditions $(iii)$ and $(iv)$  of Definition \ref{rel}.
In an analogous way we define a problematic cycle of $C_2$.

To compute a relaxed cycle cover  $C_1$  improving $C_{max}$ and  a relaxed cycle cover improving $C_{max}$ and $C_1$  we construct the following undirected  graphs $G'=(V',E')$ and  $G'' =(V'',E'')$.
 
First we describe graph $G'$.
For each vertex $v$ of $G$ we add two vertices $v_{in}, v_{out}$ to $V'$. For each edge $(u,v) \in E$
we add vertices $e^1_{uv}, e^2_{uv}$, an edge  $(e^1_{uv}, e^2_{uv})$ of weight $0$ and edges $(u_{out}, e^1_{uv}), (v_{in}, e^2_{uv})$, each of weight $\frac{1}{2}w((u,v))$.
Next we build so-called gadgets.

For each $2$-cycle of $C_{max}$  on vertices $u$ and $v$  we add vertices $a_{\{u,v\}}, b_{\{u,v\}}$ 
and edges $(a_{\{u,v\}}, e^1_{uv}), \\ (a_{\{u,v\}}, e^2_{vu}, (b_{\{u,v\}}, e^1_{vu}), (b_{\{u,v\}}, e^2_{uv})$ having weight $0$.

Graph $G''$ is an extension of graph $G'$. Let $C_1$ be any relaxed cycle cover improving $C_{max}$. Additional gadgets contained in $G''$ are as follows.

For each  triangle $t$  on vertices $p,q,r$ containing edges $(p,q), \\ (q,r), (r,p)$ such that either (1) $t \in C_{max} \cap C_1$ or (2) $t \in C_{max}$ and $C_1$ contains a $2$-cycle on two vertices of $t$, we add vertices $a_{\{p,q,r\}}, b_{\{p,q,r\}}$ and connect them appropriately to vertices $e^2_{pq}, e^1_{rq}, e^1_{rp}$ and $e^1_{pq}, e^2_{rq}, e^2_{rp}$
via edges of weight $0$. The gadget is depicted in Figure \ref{gtriangle}. Let us notice that it cannot happen that $w((p,r)) > \frac{1}{2}(w((p,q))+w((q,r))$
and $w((q,p)) > \frac{1}{2}(w((q,r))+w((r,p))$ and $w((r,q)) > \frac{1}{2}(w((p,q))+w((r,p))$ because it would mean that triangle consisting of edges $(q,p), (r,q), (p,r)$ has greater weight than $t$ and thus that $C_{max}$
is not a maximum weight cycle cover of $G$. Therefore w.l.o.g. we can assume that $w((p,r)) \leq \frac{1}{2}(w((p,q))+w((q,r))$.

For each $4$-cycle $c$  such that $c \in C_1$ and  $C_{max}$ has two $2$-cycles that share an edge with $c$ or  $c \in C_{max}$ and  $C_1$ has two $2$-cycles that share an edge with $c$, we add a similar gadget which will enforce
that at most six half-edges of edges $c \cup c'$ are present in $\tilde C$.

\begin{theorem}
Any perfect matching of $G'$ yields  a relaxed cycle cover  $C_1$ improving $C_{max}$.
Any perfect matching of $G''$ yields a relaxed cycle cover improving $C_{max}$ and $C_1$.
A maximum weight perfect matching of $G'$ yields a relaxed cycle cover $C_1$ improving $C_{max}$  such that $w(C_1) \geq OPT$.
A maximum weight perfect matching of $G''$ yields a relaxed cycle cover $C_2$ improving $C_{max}$ and $C_1$ such that $w(C_2) \geq OPT$.
\end{theorem}
\dowod  First we will show that any perfect matching of $G''$ yields  a relaxed cycle cover  improving $C_{max}$ and $C_1$.  
Let $M$ be any perfect matching of $G''$. $M$ defines a set of half-edges $\tilde C \subseteq \tilde E$ of the same weight that satisfies property $(i)$ of Definition \ref{rel}. Let us now verify that the thus defined set $\tilde C$ satifies
also the other properties of Definition \ref{rel}.

Let $c$ be a $2$-cycle of $G$ on vertices $u,v$. Then $G'$ contains vertices $a_{\{u,v\}}, b_{\{u,v\}}$ and the fact that these vertices  have to be matched in a perfect matching ensures that $\tilde C$ does not contain all four half-edges corresponding to $(u,v)$ and $(v,u)$.  If vertices $a_{\{u,v\}}, b_{\{u,v\}}$ are matched so that  $a_{\{u,v\}}$ is matched with $e^1_{uv}$ and $ b_{\{u,v\}}$  is matched with $e^2_{uv}$,  or  $a_{\{u,v\}}$ is matched with $e^2_{vu}$ and $ b_{\{u,v\}}$  is matched with $e^1_{vu}$ , then both half-edges of one of the edges $(u,v), (v,u)$ are excluded from $\tilde{C}$ and for the other edge, either both of its half-edges belong to $\tilde{C}$ or both do not belong to $\tilde C$.
If   $a_{\{u,v\}}, b_{\{u,v\}}$ are matched in the other way, then  matching $M$ does not contain appropriately either (1) $(u_{out}, e^1_{uv})$ and $(e^1_{vu}, v_{out})$
or (2) $(u_{in}, e^2_{vu})$ and $(e^2_{uv}, v_{in})$.  Therefore condition $(ii)$ is indeed satisfied.

Let $t$ be a triangle  containing edges $(p,q), (q,r), (r,p)$ satisfying the properties given in Definition \ref{rel}. Suppose that $w((r,p)) \leq \frac{1}{2}(w((p,q))+w((q,r))$. Then the gadget for $t$ in $G''$ looks as in Figure \ref{gtriangle}.
Each of the pair of vertices $a_{\{p,q\}}, b_{\{p,q\}}$ and $a_{\{r,q\}}, b_{\{r,q\}}$, and $a_{\{p,r\}}, b_{\{p,r\}}$ ensures that exactly two half-edges of edges belonging to corresponding $2$-cycles are excluded from $\tilde C$. Vertices $a_{\{p,q,r\}}$ and $b_{\{p,q,r\}}$ gurantee that two more half-edges connecting two vertices from the set $V_t$ are excluded from $\tilde C$. Therefore $\tilde C$ contains at most four half-edges 
connecting two vertices from the set $V_t$.

We will prove now that $\tilde C$  contains neither all the half-edges from the set $S_1=\{(r,q), (p, x_{(p,q)}), (x_{(r,p)},p)\}$ nor all the half-edges from the set $S_2=\{(q,p), (r, x_{(r,p)}), (x_{(q,r)},r)\}$.
Let us suppose  that $\tilde C$ contains all the half-edges from the set $S_1$. Since edge $(r,q)$ belongs to $\tilde C$ vertex  $b_{\{p,q,r\}}$  cannot be matched with 
$e^2_{(r,q)}$.  Since both $(p, x_{(p,q)})$ and $(x_{(r,p)},p)$ belong to $\tilde C$ vertex  $b_{\{p,q,r\}}$ cannot be matched with $e^1_{(p,q)}$ or with $e^2_{(r,p)}$. But $b_{\{p,q,r\}}$ is incident only with
vertices $e^2_{(r,q)}, e^1_{(p,q)}, e^2_{(r,p)}$ and these are the only vertices it can be matched with - a contradiction.

Let us now suppose that $\tilde C$ contains all the half-edges from the set $S_2$. Since edge $(q,p)$ belongs to $\tilde C$, vertex $b_{\{p,q\}}$ must be matched with  $e^2_{(p,q)}$. Since $(x_{(q,r)}, r) \in \tilde C$ vertex
$b_{\{q,r\}}$ must be matched with $e^1_{(r,q)}$. Since $(r, x_{(r,p)}) \in \tilde C$ vertex $a_{\{p,q,r\}}$ cannot be matched with $e^1_{(r,p)}$. But   $a_{\{p,q,r\}}$ is incident  only with $e^2_{(p,q)}, e^1_{(r,q)}, e^1_{(r,p)}$- a contradiction.

It is fairly straightforward to show that each traveling salesman tour $T$ of $G$  corresponds to some perfect matching of $G''$. It is so because a tour cannot contain a $2$-cycle or triangle if the graph $G$ contains at least four vertices. Therefore any maximum weight perfect matching of $G''$ yields a relaxed cycle cover $C_2$ improving $C_{max}$ and $C_1$ such that $w(C_2) \geq OPT$.

The proof for a relaxed cycle cover  $C_1$ improving $C_{max}$  is analogous.  \koniec

\begin{figure}
\centering{\includegraphics[scale=0.6]{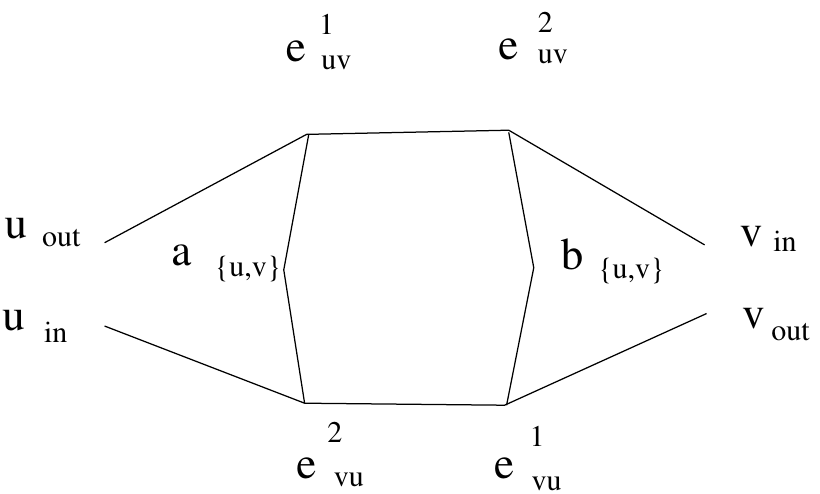}}
\caption{{\scriptsize  A gadget corresponding to a $2$-cycle  on vertices $u,v$.}
} \label{2cykl}
\end{figure}

Let us notice that a relaxed cycle cover $C_1$  and a relaxed cycle cover $C_2$ improving $C_1$ and $C_{max}$ consist either of (1) directed cycles or (2) directed paths such that each one of them begins and ends with a half-edge or both. 

We will say that $C_1$ ($C_2$) is {\bf \em integral} if for each edge $e \in E$ $C_1$ ($C_2$) contains either both half-edges of $e$ or none. 
If $C_1$ or $C_2$ is integral, then it corresponds to a cycle cover of $G$ i.e. consists only of directed cycles.

\section{$\frac 34$ - approximation algorithm for Max ATSP}

The algorithm starts with computing a maximum weight cycle cover $C_{max}$ of a given graph $G$.  If $C_{max}$ does not contain any hard cycles,  we can easily build a tour of weight $\frac{3}{4} OPT$. Otherwise we compute a relaxed cycle cover $C_1$  improving $C_{max}$  and, if necessary,  a relaxed cycle cover $C_2$ improving $C_{max}$ and $C_1$. Both of them are  such that $w(C_1), w (C_2) \geq OPT$ . We compute $C_1$ and $C_2$ in the way described in the previous section.
We create two multigraphs $G_1$ and $G_2$ as described in Introduction.
The total weight of all edges of $G_1$ (corr. $G_2$)  is equal to $w(C_{max} + 2 w(C_1)$ (corr.  2($w(C_{max})+w(C_1)+w(C_2))$), which is at leas t$3 OPT$ (corr. $6 OPT$.). Next we would like to divide edges of $G_1$ into four sets $Z_1, Z_2, \ldots, Z_4$ in such a way that each $Z_i$ ($1 \leq i \leq 4$) is a collection of
vertex-disjoint paths.  If this turns out to be impossible, because $C_1$ contains a problematic cycle, we will show how to divide edges of $G_2$ into eight sets $Z_1, Z_2, \ldots, Z_8$ in such a way that each $Z_i$ ($1 \leq i \leq 8$) is a collection of
vertex-disjoint paths.The  task  of dividing edges of $G_1$ or $G_2$ into appropriately four or eight sets can equivalently be viewed as coloring each edge of $G_1$  or $G_2$ with one of four or correspondingly one of eight colors. Let $\Ko$ denote $\{1,2,3,4,5,6,8\},$  $\Kr=\{1,2,3,4\}$ and $\Kt=\{1,2,3\}$.

Further on we will treat $G_1$ and $G_2$ as simple graphs but will require that each edge is colored with an appropriate number of colors equal to the number of its copies in the graph.

We will say that two edges $e_1$ and $e_2$ are {\bf \em in-coincident} if there exists a vertex $v$ such that both of them are going into $v$. ( We will also say that $e_1$ is in-coincident with $e_2$.) We will say that they are  {\bf \em out-coincident} if there exists a vertex $v$ such that both of them are outgoing from $v$. Two edges are {\bf \em coincident} if they are in-coincident or out-coincident. We will also say that an edge $e$ is in-coincident with a cycle $c$ of $C_1$ if $e$ is in-coincident with some edge of $c$.

Cycle covers are a special case of $b$-matchings and therefore many notions and facts from matching theory will prove useful. Below we recall some of them.
A sequence $P$ of edges $(e_1, e_2, \ldots, e_k)$ is said to be {\bf \em an alternating  path  (with respect to a given cycle cover $C$)} if its edges alternately belong and do not belong to $C$, every two consecutive edges of $P$ are coincident and for each $1 < i < k$ 
edge $e_i$ is in-coincident with $e_{i-1}$ and out-coincident with $e_{i+1}$ or vice versa.  An alternating  path $P=(e_1, e_2, \ldots, e_k)$ such that exactly one of the edges $e_1, e_k$ belongs to $C$
and either $e_k$ is in-coincident with $e_{k-1}$ and out-coincident with $e_1$ or the other way round is called {\bf \em an alternating  cycle  (with respect to $C$)}.
It can be  observed that a symmetric difference $C_1 \oplus C_2=(C_1 \setminus C_2) \cup (C_2 \setminus C_1)$ of two cycle covers $C_1, C_2$ of $G$ consists of edge-disjoint alternating cycles (wrt. both $C_1$ and $C_2$). (See also \cite{Lovasz} for example.)

Since $C_{rel}$ is not  always integral, we  need to extend the definition of an alternating cycle. We do it as follows. Let $C$ be any cycle cover of $G$.
A sequence $P=(e_1, e_2, \ldots, e_k)$ is going to be called {\bf \em an alternating h-cycle  (wrt. $C$)} if $e_1$ and $e_k$ are half-edges and otherwise $P$ satisfies analogous conditions as an alternating path.

Wel also introduce the term of an alternating weight. Suppose that $P$ is an alternating path or cycle  (wrt. $C$), then {\bf \em the alternating weight  (wrt. $C$) of $P$} is defined
as $\tilde{w}_C (P) = \sum_{e \in P \setminus C} {w(e)} - \sum_{e \in P \cap C} {w(e)}$. If $C$ is a cycle cover of $G$ and $A$ an alternating cycle  wrt. $C$, then by {\bf \em applying $A$ to $C$} we will mean the operation, whose result is $C \oplus A$.
$C \oplus A$ is also a cycle cover of $G$ and we have $w(C \oplus A)= w(C) + \tilde{w}_C(A)$. We can easily observe the following property of any alternating cycle with respect to $C_{max}$.

\begin{fact}\label{F}
Let $A$ be any alternating  cycle  with respect to $C_{max}$ . Then $\tilde{w}_{C_{max}}(A) \leq 0$.
\end{fact}
\dowod
$C_{max}$ is a maximum weight cycle cover of graph $G$ and $C_{max} \oplus A$ is a cycle cover of $G$. Therefore $w(C_{max} \oplus A) \leq w(C_{max})$. Since  $w(C \oplus A)= w(C) + \tilde{w}(A)$, we get that $\tilde{w}(A) \leq 0$.
\koniec

Both $C_{max} \oplus C_1$ and $C_{max} \oplus C_2$ disintegrate into alternating cycles and h-cycles, i.e. each edge of correspondingly   $C_{max} \oplus C_1$ and $C_{max} \oplus C_2$  belongs to exactly one alternating cycle or h-cycle. \\
We will say that an alternating cycle $A$ of  $C_{max} \oplus C_1$ is {\bf \em necessary} if $C_1 \oplus A$ contains a $2$-cycle which also belongs to $C_{max}$. \\
We will say that an alternating cycle $A$ of  $C_{max} \oplus C_2$ is {\bf \em necessary} if $C_2 \oplus A$ contains any of the cycles described in Definition \ref{rel}.

It will be  convenient for us to deal with relaxed cycle covers $C_1$ and $C_2$ such that both $C_{max} \oplus C_1$ and $C_{max} \oplus C_2$  contain only necessary alternating cycles and possibly alternating h-cycles.

\begin{lemma}\label{necessary}
Let $C_1$ be a relaxed cycle cover improving $C_{max}$ such that $w(C_1) \geq OPT$ and let $C_2$ be a relaxed cycle cover improving $C_{max}$ and $C_1$  such that $w(C_2) \geq OPT$. Then there exists a relaxed cycle cover $C'_1$ improving $C_{max}$ such that
$w(C'_1) \geq w(C_1)$ and every alternating cycle from $C_{max} \oplus C'_1$  is necessary. Similarly, there exists a relaxed cycle cover $C'_2$ improving $C_{max}$ and $C'_1$  such that
$w(C'_2) \geq w(C_2)$ and every alternating cycle from $C_{max} \oplus C'_2$  is necessary.
\end{lemma}

Because of the above lemma since this time on we will assume that every alternating cycle from $C_{max} \oplus C_1$  or  $C_{max} \oplus C_2$  is necessary.

\subsection{Coloring of $G_1$}
We will now show how to color the edges of $G_1$ with colors from $\Kr$ in such a way that each color class  consists of vertex-disjoint paths. (If it is not possible to color the whole $G_1$, then we leave problematic cycles uncolored.)

Let us first demonstate how we color the graph $G_1$ when $C_1$ is integral.

Formally, the coloring of $G_1$ is an assignment of colors of $\Kr$ to edges of $G_1$. Each edge of $C_{max}$ is assigned one color and each edge of $C_1$ is assigned two colors. (If an edge belongs to $C_{max} \cap C_1$, then ,as a consequence, it is assigned three colors.) We say that a coloring of $G_1$ is  {\bf \em partial}  if some of its edges have colors assigned to them (these edges are {\bf \em colored}) and some of them do not (these edges are {\bf \em uncolored}.)  Still, in a partial coloring of $G_1$  an edge of $C_1$ either has two colors assigned to it or none.  A  (partial) coloring  of $G_1$ is called {\bf \em good} if each color class  consists of vertex-disjoint paths

The idea behind coloring $G_1$ is that for each cycle of $C_{max}$ we mark at least one of its edges, that is going to be colored with one of the colors of $\Kt$ and color the unmarked edges of $C_{max}$  with $4$. Whenever possible we want to color the edges of $C_1$  only with colors of $\Kt$. Each edge $e=(u,v)$ of $C_1$  has to be colored twice. Let $e_1, e_2$ be two edges of $C_{max}$ that are coincident with $e$, i.e. they have the form $e_1=(u,v'), e_2=(u',v)$  (they may denote the same edge).  Then, of course $e$ cannot be colored with the same color as $e_1$ or $e_2$. If $e_1$ is marked and we want to color it $1$ and $e_2$ is unmarked, then $e$ has to be colored with $2$ and $3$.  If both $e_1$ and $e_2$ are marked and we want to color $e_1$ with $1$ and $e_2$ with $2$, then we have to use $4$ for coloring $e$ and color it with $3$ and $4$. On the other hand, if both $e_1$ and $e_2$ are marked and we want to color them with the same color, say $1$ for example, then we have a choice and can color $e$ with $2$ and $3$ or $2$ and $4$ or $3$ and $4$.   

In a (partially) colored  $G_1$ a cycle $c$  is called {\bf \em monochromatic} if there exists a color $i$ of $\Kr$ such that each edge of $c$ is colored $i$.  Of course, if a (partial) coloring of $G_1$ is good, then $G_1$ does not contain any monochromatic cycles.
We will say that an edge $e$ is {\bf \em safe} if  no matter how we color the so far uncolored edges, it is guranteed not to belong to any monochromatic cycle. For example suppose that edge $e=(u,v) \in C_{max}$ is colored with color $1$ and that edges $(z,u) \in C_{max}, (z',u) \in C_1$ are colored appropriately $2$ and $3,4$. Then clearly $e$ is safe. An edge $e=(u,v) \in C_{max}$ will be called {\bf \em external} if $u$ and $v$ belong to two different cycles of $C_1$. Otherwise it will be called {\bf \em internal}.

We state now the  observation that will prove very useful in guaranteeing that we do not create monochromatic cycles.
\begin{observation} \label{obs}
Suppose that edge $e=(u,v)$  colored $1$ goes into cycle $c$ of $C_1$ (i.e. $u$ does not belong to $c$ and $v$ does).
If there exists no edge $e'=(u',v')$ going out of $c$ ($u'$ is on $c$ and $v'$ is not) that is colored $k$, then $e$ does not belong to any cycle, whose all edges are colored $k$. In other words, $e$ is safe.
\end{observation}

We say that an edge $e$ belonging to a cycle of $C_1$ is  {\bf \em winged} if all external edges coincident with it are marked. (An edge that has no external edges coincident with it is thus wingy as well.)
For each cycle $c$ of $C_1$ we will require that at least $|c|-1$  of its edges are winged. ($|c|$ denotes the length of $c$.)  An external marked edge $e$  coincident with a cycle $c$ is called {\bf \em a tail (of $c$)} if all external edges coincident with $c$ are marked or if  $e$ is coincident with an edge of $c$
that is not winged. An external marked edge $e$ coincident with a cycle $c$ that is not its tail is a {\bf \em wing (of $c$)}. We also say that a cycle $c$ is {\bf \em taily} if $c$ has exactly $2|c|$ external marked edges coincident with it.

Now, we are ready to give the definition of a nice set of marked edges.
\begin{definition} \label{nice}
We say that a set  of marked edges of $C_{max}$ is {\bf \em nice} if
\begin{itemize}
\item for each cycle $c$ of $C_1$ at least $|c|-1$ of its edges are winged,
\item no edge of $C_{max}$ is a tail of two different cycles of $C_1$,
\item for each $2$-cycle $c$ of $C_1$, if $c_1$ has exactly one tail, then it also has a wing which is also a wing of another cycle of $C_1$.  
\end{itemize}
\end{definition}

\begin{lemma} \label{mark}
There exists an algorithm of marking the external edges of $C_{max}$  in such a way that the resulting set is nice. The running time of the algorithm is $O(n)$.
\end{lemma}

Next we are going to show the algorithm of coloring $G_1$ which consists of three phases. In Phase 1 we color all marked external edges in such a way that all of them are safe. 
For clarity, we first present Phase 1 and Phase 2 for the case when there are no taily cycles.\\

{\scriptsize
\noindent Phase 1 of Algorithm Color $G_1$ \\
\noindent {\bf while} there exists a cycle $c$ of $C_1$ that has an uncolored tail or wing {\bf do} \\
\indent \indent  color all the uncol.  marked external  edges going into $c$ with $1$ and all the uncol.  marked external edges going out of $c$ with $2$ or $3$; \\
}

\begin{lemma}\label{safe}
In Phase 1 of Algorithm Color $G_1$, after getting colored, a marked external edge $e$ is safe.
\end{lemma}  
\dowod
Let $e$ denote any external marked edge.
 At some point, $e$ gets colored together with all other so far uncolored external marked edges coincident with some cycle $c$ of $C_1$.  If at this point, $c$ has no previously colored marked external edges coincident with it, then by Observation \ref{obs}, $e$ becomes safe, because we color all marked edges going into $c$ with $1$ and all marked edges going out of $c$ with $2$ or $3$. If, on the other hand, $c$ has some previously colored marked external edges coincident with it, then, by this proof, they are already safe. Thus, if $e$ is going into $c$, then we color it with $1$ and if there exists an edge $e$ going out of $c$ and colored $1$, then it must have been colored previously and is safe. Therefore, $e$ is safe as well. The case when $e$ is going out of $c$ is analogous. \koniec

After Phase 1 all colored edges are safe but it may happen that there is no way of completing this coloring so that it is good. For example if a cycle $c$ has $2c$ external edges coincident with it and all edges going into $c$ are colored $1$ and all edges going out of $c$ are colored $2$, then the only way of completing the current coloring is to color each edge of $c$ with $3$ and $4$. This way we would create a cycle within the class of edges colored $3$ and within the class of edges colored $4$. We say that a cycle $c'$ of $G_1$
is a  {\bf \em cycle within a cycle $c$} if every vertex lying on $c'$ lies also on $c$.
We say that a cycle $c$ of $C_1$ is blocked if there is no way of completing the current partial coloring so that it does not contain a monochromatic cycle within $c$. An internal edge $e=(u,v)$ will be called {\bf \em quasiexternal} if there exist external edges $e_1, e_2$ such that $e_1$ is incident with $u$ and $e_2$ is incident with $v$.

\begin{lemma}\label{block}
A cycle $c$ of $C_1$ can be blocked only if $c$ satisfies one of the following:
\begin{itemize}
\item $c$ has exactly $2|c|$  external  edges coincident with it and the number of edges of $c$  whose wings are colored the same plus the number of colors of $\Kr$  that occur on the external edges of $c$ is less than $4$,
\item all external edges of $c$ are colored with the same color,
\item all external edges coincident with $c$ are marked, all internal edges of $c$ are quasiexternal and for each quasiexternal edge $e=(u,v)$ of $c$   there exist two different colors $i_1, i_2$ of $\Kt$ such that an external edge of $c$ incident with $u$ is colored $i_1$,
an external edge of $c$ incident with $v$ is colored $i_2$ and an edge of $c$ that is coincident with $e$ has an external marked edge coincident with it that is colored with a color $i_3$ belonging to $\Kt \setminus \{i_1, i_2\}$,
\item $c$ is problematic.
\end{itemize}
Otherwise the current partial coloring can be completed in such a way that it does not contain  a monochromatic cycle within $c$.
\end{lemma} 

Let us notice that if a cycle $c$ of $C_1$ is blocked and it is not taily, then we can change the coloring of its tail  $t$ to $k$, where $k$ is such a color of $\Kt$ that no external edge incident with $c$ is colored $k$, and, as a result, $c$ will not be  blocked any more and $t$ will be safe. Roughly, this is what we are going to do in Phase 2 with blocked cycles that are not taily.

In Phase 2 we are going to change the coloring of some of the external marked edges in such a way that each colored edge is safe and there is no blocked cycle of $C_1$. \\

{\scriptsize
\noindent Phase 2 of Algorithm Color $G_2$ \\

\vspace{-0.2cm} {\bf while} there exists a blocked cycle of $C_1$ that is not taily \\

\vspace{-0.2cm} \hspace{2cm} $c \leftarrow$ any blocked cycle of $C_1$  that is not taily\\

\vspace{-0.2cm}\hspace{2cm}{\bf  while}  $c$ is blocked {\bf do} \\

\vspace{-0.2cm}\hspace{3cm} change the coloring of the tail $t$  of $c$  so that $c$ is not blocked any more \\

\vspace{-0.2cm}\hspace{3cm} $t$  is also a wing of another cycle $c'$ of $C_1$ \\

\vspace{-0.2cm}\hspace{3cm} {\bf if}  $c'$ is blocked and is a $2$-cycle, {\bf then} we can change the coloring of its wings and a tail in such a way\\

\vspace{-0.2cm}\hspace{3cm}  that we do not create a new blocked cycle \\

\vspace{-0.2cm}\hspace{3cm} {\bf else  if} $c'$ is blocked, {\bf then} $c \leftarrow c'$

}

\begin{lemma} \label{P2}
In Phase 2 of Algorithm Color $G_1$ each colored edge is safe at all times. After executing Phase 2, the graph $G_1$ does not contain a blocked cycle $c$ of $C_1$, apart from problematic cycles. The running time of Phase 2 is $O(n^2)$.
\end{lemma}
\dowod
Let $G_s$ denote the graph $G_c$ in which every cycle of $C_1$ is shrunk into one vertex. All internal edges become loops in $G_s$ and we remove them (from $G_s$). Thus, the edges of $G_s$ contain all external edges. Let $ G_s^j$ denote the subgraph of $G_s$ containing  edges colored $j$.  Let us observe that  saying  that  saying that graph $G_s^j$ is acyclic  implies that all external marked edges colored $j$ are safe. Before starting Phase 2 each graph $G_s^j$ for $j \in \{1,2,3\}$ is acyclic by Lemma \ref{safe} and the properties of Phase 1. When in Phase 2 of Algorithm Color $G_1$  we change the coloring of edge $e$ from $i_1$ to $i_2$, we remove one edge from graph $G_s^{i_1}$ and add one edge to graph $G_s^{i_2}$.  Removing an edge from an acyclic graph cannot make it non-acyclic. When we change the coloring of $e$ into $i_2$,
then one of the cycles of $C_1$ with which $e$ is coincident has no other external edges coincident with it that are colored $i_2$. Therefore adding such an edge to $G_s^{i_2}$ cannot create a cycle there either.

If a cycle $c$ is blocked, we change the coloring of its tail. Another time the coloring of some external edge coincident with $c$ can change is when one of its wings changed a color. Then if $c$ is not a $2$-cycle, it cannot become blocked and if $c$ is a $2$-cycle,
we are able to change the coloring of its wings and tail in such a way that we do not craete a new blocked cycle. This shows that the inner while loop in Phase 1 ends after visiting at most all cycles of $C_1$. \koniec

Now, we show what to do with taily cycles. We say that a taily cycle $c$  is {\bf \em  favourable} if   there exitsts such an edge of $c$ that the two tails of $c$ coincident with $e$ are wings of two different cycles
of $C_1$.  For each taily favourable cycle $c$ of $C_1$ that is not a $2$-cycle, we will be able to guarantee that it does not become blocked in Phase 1 and for each taily cycle $c$ that is not favourable we will be able to guarantee that for each color $k $ of $\Kt$
at least one of the tails of $c$  is colored $k$. 

{\scriptsize
\noindent Phase 1 of Algorithm Color $G_1$ \\
\noindent {\bf while} there exists a cycle $c$ of $C_1$ that has an uncolored tail or wing {\bf do} \\
\indent \indent {\bf while} there exists a taily favourable cycle $c$ that has an edge with exactly one colored tail coincident with it \\ 
\indent \indent \indent let $e$ be an edge of $c$ such that exactly one of the edges $e_1, e_2$ coincident with it is colored; \\
\indent \indent \indent suppose that $e_1$ is colored $k$ and is going into (out of)  $c$; \\
\indent \indent \indent color all the uncol. external  edges going into (out of)  $c$ with $k$ and all the uncol. external edges going out of (into) $c$ \\
\indent \indent \indent  with one of the colors  from $\Kt \setminus \{k\}$ .\\
\\
\indent \indent  {\bf while} there exists a taily  cycle $c$ that is not favourable and whose at least one tail is colored \\ 
\indent \indent \indent suppose that one of the tails of $c$ is colored $k$; \\
\indent \indent \indent let $k_1,k_2$ be the two colors from  $\Kt \setminus \{k\}$; \\
\indent \indent \indent   color all the uncol. external  edges going into $c$ with $k_1$ and all the uncol. external edges going out of $c$ with $k_2$ or vice versa\\
\indent \indent \indent so that $c$ has a tail colored with every color from $\Kt$. \\
\\
\indent \indent  color all the uncol.  marked external  edges going into $c$ with $1$ and all the uncol.  marked external edges going out of $c$\\
\indent \indent \indent with $2$ or $3$. \\
}

\begin{lemma}\label{safe1}
In Phase 1 of Algorithm Color $G_1$, after getting colored, a marked external edge $e$ is safe. After the execution of Phase 1  no taily favourable cycle of $C_1$, which is not a $2$-cycle, is blocked.
\end{lemma}  
\dowod
The proof of the first part of the lemma goes through analogously as the proof of Lemma \ref{safe}. The key observation is that at the step when we color all uncolored external edges incident with a given cycle $c$ of $C_1$, we color all the so far uncolored edges going into $c$ 
differently than the so far uncolored edges going out of $c$.

\koniec

{\scriptsize
\noindent Part 2 of Phase 2  of Algorithm Color $G_1$ \\
\noindent {\bf while} there exists a blocked taily cycle $c$ of $C_1$  {\bf do} \\
\indent \indent if $c$ is not a $2$-cycle, then $c$ has an edge such that if we change the color of one of the tails $t$ of $c$ coincident with $e$ to $4$, \\
\indent \indent then $c$ will cease to be blocked and a cycle $c_1$ of $C_1$ such that $t$  is its wing will not become blocked; \\
\indent \indent if $c$ is a $2$-cycle, then we can change the coloring of its tails in such a way that $c$ is not blocked any more \\
\indent \indent  and we do not create a new blocked cycle.

}

\begin{lemma} \label{taily}
Part 2 of Phase 2 of Algorithm Color $G_1$ can be performed. Moreover, it can be performed in such a way that whenever an external edge changes color it becomes safe.
\end{lemma}

Phase 3 of Algorithm Color $G_1$ is based on Lemma \ref{block} and consists in completing the coloring in such a way that graph $G_1$  does not contain  a monochromatic cycle within any cycle of $C_1$ and coloring all unmarked external edges $4$.

We can observe that after the execution of Phase 3,  $G_1$ does not contain a monochromatic cycle colored $1, 2$ or $3$.  It is so, because by Lemma \ref{P2}, each external marked edge is safe and by Lemma \ref{block}, there is no  monochromatic cycle within any cycle of $C_1$. We will now prove that $G_1$  cannot contain   a monochromatic cycle colored $4$ either.  Let us notice that an edge $e \notin C_{max}$ can be colored $4$ only if it belongs to $C_1$ and two edges of $C_{max}$ coincident with it are marked.   Therefore, the danger of creating a  monochromatic cycle colored $4$ exists on a cycle $c$ of $G_1$ if every edge $e$ of $c$ is either an unmarked edge of $C_{max}$  or has two marked edges  coincident with it and $c$ contains an external edge.  Let us call such a cycle {\bf \em black}.

\begin{lemma} \label{black}
After the execution of Phase 3, the graph $G_1$ contains no black cycle.
\end{lemma}
\dowod The existence of a black cycle in $G_1$ would imply the existence of an alternating cycle $A$ of $C_{max} \oplus C_1$, which is not necessary. \koniec

\subsection{Coloring of $G_2$}
The first part of coloring $G_2$  consists in coloring $G'_2$ which consists of one copy of $C_{max}$ and two copies of $C_2$. We color $G'_2$ with colors from $\Kr'=\{5,6,7,8\}$  in a very similar way to coloring $G_1$. (In $G'_2$ directed paths may end with half-edges
of edges of a triangle or  a $4$-cycle of $C_{max}$.)  It may happen that we cannot color the whole $G'_2$ because it contains problematic cycle(s). Still, problematic cycles do not occur on the same set of vertices both in $G_1$ and $G'_2$.
As a final step, we complete the coloring of $G_1$ and $G'_2$ by 'borrowing' colors from one another. 

For example, suppose that a triangle $t$ belongs both to $C_{max}$ and $C_1$ but none of its edges belongs to $C_2$. Notice  that $t$ needs to be colored four times in $G_2$. Then, if four edges of $t$  are colored with colors from $\Kr'$, then the remaining eight edges
of $t$ can be colored with colors from $\Kr$. \\

\section{Missing proofs}

\subsection{Proof of Lemma \ref{necessary}}

\dowod 
$C_1 \oplus C_{max}$ can be decomposed into alternating cycles and h-cycles. Let $A_1, \ldots, A_k$ denote all alternating cycles from $C_{max} \oplus C_1$ that are not necessary. Then $C'_1=C_1 \oplus \bigcup_{i=1}^k A_i$ is another relaxed cycle cover improving $C_{max}$.  Moreover $C'_1 \oplus C_{max}$ does not contain any alternating cycle that is not necessary. The weight of $C'_1$ is equal to $w(C_1)+\sum_{i=1}^k \tilde{w}(A_i)$. By Fact \ref{F} the alternating (with respect to $C_{max}$) weight of $A_i$ is nonpositive.
Therefore the alternating with respect to $C_1$ weight of   $A_i$ is nonnegative. This means that $w(C'_1) \geq w(C_1)$. The proof of the existence of $C'_2$ is analogous. \koniec

\subsection{Proof of Lemma \ref{block}}

\dowod
We prove that if a cycle $c$ of $C_1$ does not fall into any of the described  five categories, then the current partial coloring can be completed in such a way that it does not contain a monochromatic cycle within $c$.

{\bf Case 1: $c$ has no external edges.} \\
 For each cycle $c'$ of $C_{max}$ whose all edges are internal edges of $c$, we choose one of its edges and color it $1$. We do it in such a way that if $c'$ contains an edge belonging also to $c$, then we color such an edge with $1$.
If possible, we choose an edge $e$ of $c$ such that the edges $e_1, e_2$ of $C_{max}$ belong to two different cycles of $C_{max}$ and color $e_1$ and $e_2$ with $1$ and $e$ with $2$ and $4$. We then find one more edge of $c$ that can be colored with $1$ and $3$ and we are done.

{\bf Case 2: $c$ has external edges}\\
For each edge $e$ of $c$ that has one colored edge $e'$ coincident with it, we color $e$ with colors belonging to $\Kt \setminus k$, where $k$ is the color with which $e$ is colored.
For each edge $e$ of $c$ that has two differently colored edges $e_1, e_2$ coincident with it, we color $e$ with colors from $\Kr \setminus \{k_1,k_2\}$, where $k_1, k_2$ are the colors with which $e_1, e_2$ are colored.
If edge $e$ of $c$ that has two colored edges $e_1, e_2$ coincident with it that are colored with the same color $k \in \Kt$, then if for each color $k \in \Kt$ there exists an edge $e \in C_{max}$ incident with $c$ that is colored $k$, then we color $e$ with $\Kt \setminus k$, otherwise we color it with $\Kr \setminus \{k,k'\}$, where $k'$ is that color of $\Kt$ with which no edge of $C_{max}$ coincident with $c$ is colored.

If there exists a quasiexternal edge $e=(u,v)$ such that coloring it with $4$ would result in creating a monochromatic cycle within $c$, then we color it with the color with which $e_1$ or $e_2$ ic colored, where $e_1, e_2$ are external edges incident appropriately with $u$ or $v$. We try to do this in such a way that an edge $e'$ of $c$ coincident with $e$ does not have two differently colored edges coincident with it.

Similarly, if there exists a path consisting of internal and uncolored edges of $c$ such that if we colored all of the edges on this path with $4$, we would create a monochromatic cycle within $c$, we color one of these edges with a color of $\Kt$.

If a colored edge $e$ of $c$ is such that at some point a so far uncolored edge coincident with it gets colored, then we may have to change the coloring of $e$ in the way described above.

If there is no cycle of $C_{max}$, that consists solely of uncolored and internal edges of $c$, then we complete the coloring by coloring the uncolored internal edges of $c$ with $4$ and the uncolored edges of $c$ with colors from $\Kt$.  Otherwise we continue the coloring in the way shown below.
 
{\bf Case 2a: For each color $k \in \Kt$ there exists an edge $e \in C_{max}$ incident with $c$ that is colored $k$ }\\  

We proceed as follows. Let $c'$ be any cycle of $C_{max}$, whose all edges are internal edges of $c$. Let $e=(u,v)$ be any edge of $c'$. We can notice that it is always possible to color $e$ with one of the colors of $\Kt$ so that $e$ does not belong to a monochromatic cycle within $c$. It is so because an edge $e_3$ of $c$ going into $u$ is colored with at most two colors of $\Kt$.  It may happen that coloring $e$ with such a color forces  edges $e_1, e_2$  of $c$ coincident with $e$ to be colored with $4$ and as a result we could create a monochromatic cycle colored $4$ that contains  the uncolored edges of $c'$. It is possible only if both $e_1$ and $e_2$ has an edge coincident with it that is colored with a color of $\Kt$  different from $k$. Let $e'=(v',u)$ be an edge of $c'$ and let $e_4=(v,u')$ be an edge of $c$. \\
{\bf Case  i} Both  $e_3$ and $e_4$ has an edge coincident with it that is colored $k$ and $e_1$ has an edge coincident with it that is colored $k' \neq k$. Then we color $e$ with $k'$, $e'$ with $k$, $e_1$ and $e_3$ with $\Kr \setminus \{k,k'\}$ and $e_4$ with $\Kt \setminus k$.  We color all the remaining uncolored edges of $c'$ with $4$. This way, all the edges of $c'$ and edges $e_1, e_3, e_4$  are safe. \\
{\bf Case ii}    $e_3$ (or  $e_4$)  has an edge coincident with it that is colored $k$ and $e_1$ has an edge coincident with it that  is also  colored  $k$. Then we color $e$ with $k$ and $e_1$ and $e_3$ (or $e_1$ and $e_4$) with $\Kt \setminus k$. \\

{\bf Case 2b: There exists a color $k'' \in \Kt$ such that  exists no edge $e \in C_{max}$ incident with $c$  is colored $k''$ }\\  

Each cycle of $C_{max}$  but one whose all edges are internal edges of $c$  is processed as in the case 2a above.
Let $c'$ be the last cycle of $C_{max}$ such that all its edges are so far uncolored and are internal edges of $c$. Let $e=(u,v), e'=(v',u)$  be any two edges of $c'$, $e_1, e_2$ two edges of $c$ coincident with $e$ and $e_3=(v_2, u), e_4=(v, u_2)$ edges of $c$. \\
{\bf Case i}  Both  $e_1$ and $e_2$ has an edge coincident with it that is colored $k$ and $e_1$ has an edge coincident with it that is also  colored $ k$. Then we color $e$ with $k$, $e_3$ and $e_2$ with $\Kt \setminus k$ and $e_1$ with $\Kr \setminus \{k,k''\}$. 
This way we have guaranteed that at least one edge of $c$ is nor colored with $k''$. \\
{\bf Case ii} There is no edge $e$ of $c'$ that fullfills the conditions of case i above and $c'$ is a $2$-cycle. Then either (1)  both $e_1$ and $e_2$  has an edge coincident with it that is colored $k$ and $e_3$ and $e_4$ has an edge coincident with it that is colored $k' \neq k$
or (2) edges coincident with $e_1$ and $e_2$ are colored differently (with colors $k$ and $k'$) and the same for edges coincident with $e_3$ and $e_4$.  In the first case  we color $e$ with $k$, $e'$ with $k'$, $e_1$ and $e_3$ with $\Kr \setminus \{k,k'\}$, $e_2$  with $\Kt \setminus k$ and $e_4$ with $\Kr \setminus \{k',k''\}$. In the second case  we color $e$ with $k$, $e'$ with $k'$, three of the edges of $e_1, e_2, e_3, e_4$ with $\Kr \setminus \{k,k'\}$ and  one of them appropriately  with $\Kr \setminus \{k,k''\}$ or with $\Kr \setminus \{k',k''\}$. \\
{\bf Case iii}   There is no edge $e$ of $c'$ that fullfills the conditions of case i above and $c'$ has more than two edges. We consider three edges $e, e', e''$ of cycle $c'$ and proceed similarly as above ensuring that there exists an edge colored with  $\Kr \setminus \{k',k''\}$
or with  $\Kr \setminus \{k,k''\}$  and an edge of $c$ incident with $e,e'$ or $e''$ not colored with $4$.
\koniec

\subsection{Proof of Lemma \ref{taily}}
 
\dowod
Let $c$ be a blocked taily cycle of $C_1$ that has more than $2$-edges. Then for each color $k$ of $\Kt$ there exists a tail of $c$ colored $k$. Let $e$ be any edge of $c$ and $e_1, e_2$ tails of $c$ coincident with $e$.
Since $c$ is not favourable, we know that there exists a cycle $c'$ of $C_1$ such that both $e_1$ and $e_2$ are wings of $c'$. Moreover, $e_1, e_2$ are incident to two different edges $e_3, e_4$ of $c'$. (It is so because otherwise
$C_{max} \oplus C_1$ would contain an alternating cycle (consisting of four edges) that is not necessary.)  Let us notice that if $c'$ does not have a tail, then it cannot become blocked if we color $e_1$  or $e_2$ with $4$. Thus $c'$ has $2|c'|$ external edges
incident with it and is not a $2$-cycle. If changing the color of $e_1$ to $4$ causes $c'$ to become blocked and the same for $e_2$, then it means that after coloring $e_1$ with $4$, the wings of $e_1$ become colored diffeently or the number of colors of $\Kr$ occurring on the external edges incident with $c'$ descreases. In such a situation we can change the coloring of the tail $t$ of $c'$.  The tail $t$ is incident to some edge of $c'$ and some edge $e'$ belonging to another cycle $c''$. $e'$ is colored with two colors $k_1, k_2$ of $\Kr$. Therefore $t$ can be colored with the color of $\Kr$ with which it is now colored or another color $k'$ belonging to $\Kr \setminus \{k_1, k_2\}$.  Let us notice that coloring $t$ with $k'$ will ensure that $c'$ is not blocked. Also,  after coloring  $t$ with $k'$, $t$ is safe, because either $c'$
has no other external edges incident with it that are colored $k'$ or $k'=4$, which is also fine. 

\koniec

\section{Coloring of $G_1$, when $C_1$ is not integral}

\begin{figure}[ht]
\includegraphics[scale=0.7]{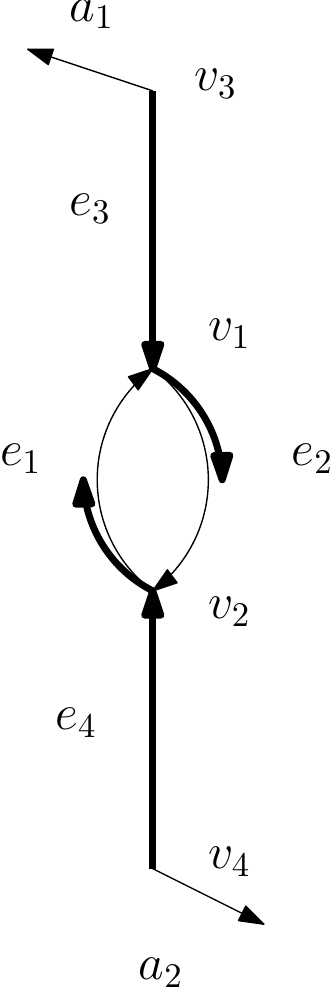}
\caption{{\scriptsize A $2$-cycle$c$  on vertices $v_1, v_2$ belongs to $C_{max}$.  $a_1$ and $a_2$ are the two antennas of $c$. Thick (half-)edges belong to $C_1$, thin ones to $C_{max}$.}
} \label{antenna}
\end{figure}

When $C_1$ is not integral, it contains half-edges. Half-edges can occur in $C_1$ only on the edges that form $2$-cycles of $C_{max}$. Moreover, if $C_1$ contains a half-edge $h_1$ of edge $e=(u,v)$, then $e$ belongs to a $2$-cycle of $C_{max}$ and $C_1$ contains also a half-edge $h_2$ of $e'=(v,u)$ and one of these half-edges is incident with $u$ and the other with $v$. Thus, either $h_1=(u, x_{(u,v)})$ and $h_2=(v, x_{(v,u)})$, or $h_1=(x_{(u,v)},v)$ and
$h_2=(x_{(v,u)},u)$.  We call edges $(u,v)$ and $(v,u)$ {\bf halfy}. Suppose that $h_1=(u, x_{(u,v)})$. Then $C_{max} \oplus C_1$ contains an alternating h-cycle of the form $(h'_1=(x_{(u,v)},v), (v_1, v), (v_1, v_2), \ldots, h_3=(v_k, x_{(v_k, v_{k+1})}))$. (Notice that $h_1 \in
C_{max} \cap C_1$ and $(u,v) \in C_{max}$, hence $h'_1 \in C_{max} \oplus C_1$.) Here an edge $(v_1, v_2)$ (belonging to $C_{max}$) is called an {\bf \em antenna of $e=(u,v)$}. If an alternating h-cycle containing $h'_1$ has the form $(h'_1=(x_{(u,v)},v), (v_1, v), h_3=(v_1, x_{(v_1, v_2)}))$, then edge $e=(v_1, v_2)$ is an antenna of  $e=(u,v)$. We say that an edge of $C_{max}$ is an antenna of a $2$-cycle $c$ of $C_{max}$, if it is an antenna of one of the edges that form $c$. 
Antennas of a $2$-cycle  are illustrated in Figure \ref{antenna}.

We can notice that in a good coloring of $G_1$  one of the edges $(u,v)$ or $(v,u)$ will have to be colored with two colors of $\Kt$.  Suppose that it is edge $(u,v)$ that we want to color with two colors of $\Kt$.  Since edge $(v_1,v)$ of $C_1$ must be colored with two colors, we will have to use $4$ for coloring it and hence an edge of $C_{max}$ that goes out of $v_1$ will have to be colored with a color different from $4$. In other words, if we want to color edge $(u,v)$ with two colors of $\Kt$, then its antenna must also be colored with a color of
$\Kt$. Since we have previously assumed that only marked edges from $C_{max}$ get colored with a color of $\Kt$, it means that if we mark one of the edges of a $2$-cycle of $C_{max}$, then we also mark its antenna. Therefore, as long as at least one of the antennas of 
a $2$-cycle $c \in C_{max}$ is marked, then we can also mark an appropriate edge of $c$ -- the one whose  antenna is marked.

When $C_1$ is not integral, it contains at least two directed paths. Each directed path of $C_1$ ends and begins with a half-edge. We extend the terminology given before as follows.
We say that an edge $e=(u,v)$ of $C_{max}$ is {\bf external} if $e$ is not halfy and  $u$ and $v$ belong to two different cycles of $C_1$  or two different paths of $C_1$ or one of the vertices $u,v$ belongs to a path of $C_1$ and the other to a cycle of $C_1$.
An edge of $C_{max}$  which is not external and is not halfy, is called {\bf internal}.

Again we mark a certain subset of external edges.

The extension of the definition of a nice set of marked edges is the following.

\begin{definition} \label{nicec1}
We say that a set  of marked edges of $C_{max}$ is {\bf \em nice} if
\begin{itemize}
\item for each cycle $c$ of $C_1$ at least $|c|-1$ of its edges are winged,
\item no edge of $C_{max}$ is a tail of two different cycles of $C_1$,
\item for each $2$-cycle $c$ of $C_1$, if $c_1$ has exactly one tail, then it also has a wing which is also a wing of another cycle of $C_1$,
\item at least one antenna of each $2$-cycle is marked,
\item for each path $p$ of $C_1$ at least $|p|-1$ of its edges are winged (by $|p|$ we mean the number of whole edges contained in $p$).
\end{itemize}
\end{definition}

\begin{lemma} \label{markc1}
There exists an algorithm of marking the external edges of $C_{max}$  in such a way that the resulting set is nice. The running time of the algorithm is $O(m)$.
\end{lemma}

We can notice the following simple fact about the possibility of a path of $C_1$ to be blocked, where we say that a path $p$ of $C_1$ is blocked if there is no way of completing the current coloring in such a way that we do not craete a monochromatuc cycle within $p$.
\begin{fact}
If in the current coloring no edge of a path $p$ of $C_1$ is colored  and  none of its internal edges, then  $p$ is not blocked.
\end{fact}

\begin{lemma}
There exists an algorithm of coloring the marked external edges such that after its execution, each colored edge is safe, no cycle of $C_1$ is blocked and two antennas of the same $2$-cycle $c$ of $C_1$ are colored differently.
\end{lemma}
This algorithm is almost the same as the one presented for the integral case.

\section{Coloring of $G_2$}
It remains to show

\begin{lemma}

Coloring of $G_2$ can be completed on problematic subgraphs.
\end{lemma}
\dowod

Let us begin with  a problematic triangle. All other cases are similar. (Moreover, it is not hard to write a computer program that would check the correctness of this lemma.)
 
If $t$ is a problematic triangle and $C_2$ contains some half-edges of edges of $t$, then we can distinguish the following cases:
\begin{itemize}
\item there are four half-edges (two ones are contained in the edges of a $2$-cycle such that one of its edges belongs to $t$, the other two share a vertex belonging to the remaining third vertex of $t$),
\item the two half-edges share a vertex and are both oriented as the edges of $t$ that contain them
\item the two half-edges share a vertex and are both oppositely oriented to the edges of $t$ that contain them
\item the common vertex of the edges containing the half-edges belongs to none of the half-edge and none, one or two of the edges containg these half-edges are oppositely oriented to the appropriate edges of $t$,
\item the common vertex of the edges containing the half-edges belongs to exactly one half-edge and none, one or two of the egses containing the half-edges   are oppositely oriented to the appropriate edges of $t$.
\end{itemize}

\begin{figure}
\centering{\includegraphics[scale=0.6]{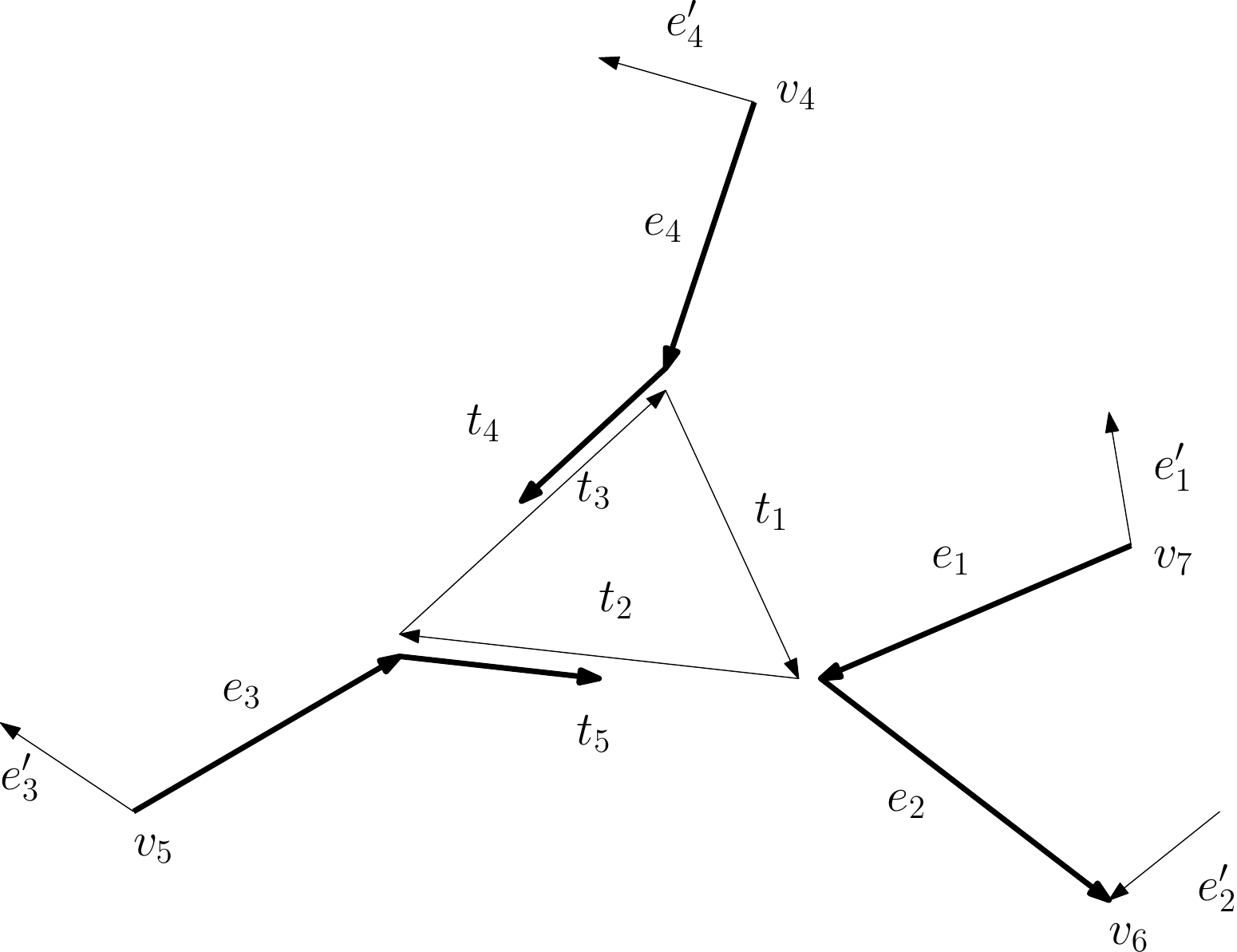}}
\caption{{\scriptsize  A triangle $t$ contains edges $t_1, t_2, t_3$. Edges $t_4$ and $t_5$ contain half-edges belonging to $C_1$. Thick (half-)edges belong to $C_1$, thin ones to $C_{max}$.}
} \label{trianglea}
\end{figure}

Let us consider the case when the common vertex of the edges containing the half-edges belongs to exactly one half-edge and both of them are oppositely oriented to the edges of $t$.
Suppose that $t$ consists of edges $t_1, t_2, t_3$,  the half-edges are contained in edges $t_4, t_5$, $e_1, e_2, e_4, e_4$  are the four egdes of $C_2$ incident with vertices of $t$ and $e'_1, e'_2, e'_3, e'_4$  are edges of $C_{max}$ coincident with edges $e_1, \ldots, e_4$. The situation is depicted in Figure \ref{trianglea}. Let us first assume that $e'_4$ is the only edge of $G'_2$  that is coincident with $e_4$ and incident with vertex $v_4$. Similarly let us assume that $e'_3$   is the only edge of $G'_2$  that is coincident with $e_3$ and incident with vertex $v_5$. Let $e_1$ be colored with colors $a$ and $b$ from $\Kr'$,  let $e'_4$ be colored with $a'$  and $e'_5$ with $b'$. Let $\{c,d\}$ denote $\Kr' \setminus \{a,b\}$.

{\bf Case 1:  It is not true that both $e_1$ and $e_2$  are colored with $b'$.}\\
If $b' \in \{c,d\}$, then  we color $t_4$ and $t_5$ with $b'$ and $t_1$ with the other color from $\{c,d\}$. Otherwise we color $t_4$ and $t_5$ with $c$ and $t_1$ with $d$.\\
If $b' \notin \{c,d\}$, then we color $t_2$ with $b'$. Otherwise we color $t_2$ with $a$ or $b$ depending on how $e_2$ is colored.  If $e_2$ is colored with $a$ and $b$, then we change its coloring so that it is colored with $a$
and one of the colors of $\Kr$ (or with $a$
and one of the colors of $\Kr$).  \\
If both $e_4$ and $e_2$ are colored with $c, d$, then we color $t_3$ with $a'$, $e_4$ with $c$ and a color belonging to $\{a,b\} \setminus \{a'\}$ and change the coloring of $e_2$ so that it is colored with $d$ and one of the colors of $\Kr$.
Otherwise we color $t_3$ with the same color as $t_2$, color $e_4$  with the appropriate two colors (i.e. not containing $a'$ or the color of $t_3$) and if $e_2$ and $e_4$ share one of the colors of $\{c,d\}$, then we remove that color from the coloring of $e_2$ and replace it with one of the colors of $\Kr$.

Let us notice that all the edges $t_1, \ldots, t_5, e_1, \ldots, e_4$  are safe and are colored an appropriate number of times as regards graph $G'_2$. However graph $G_1$ contains three copies of triangle $t$ and they cannot be colored with colors of $\Kr$. If, however,  we manage to color one of the edges of $t$ with a color of $\Kr'$, then the rest of the edges of the three copies of $t$ can be colored with colors of $\Kr$.  

When we want to replace one of the colors of $e_2$ with a color of $\Kr$, then it can be done if $G_1$ does not contain four edges going into $v_6$. $G_1$ contains four edges going into $v_6$ only if $G_1$ contains a $2$-cycle going through $v_6$, a half-edge going out of $v_6$ and an edge of $C_2$  going into $v_6$.  Such a situation can be avoided by swapping the edges and half-edges between $C_1$ and $C_2$.  Therefore we will assume that such a situation does not arise at $v_6, v_5$ or $v_4$.

Now, to be able to color the three copies of $t$ with colors of $\Kr$, we need to color one of the edges of $t$ with a color of $\Kr'$ and this must be done in addition to the coloring we have already conducted, i.e. one of the edges of $t$ should be colored with two colors
of $\Kr'$.  We can notice that we can color additionally $t_2$ or $t_3$ with $a$ or $b$. Which one we choose depends on the following. Let us assume that $e_2$ is colored with a color $k_1$ of $\Kr$.  In $G_1$ each of the vertices $v_5$ and $v_4$ has three edges going out of them, meaning that both $e_4$ and $e_5$ can be colored with remaining of the colors of $\Kr$. It may happen that that remaining color of $\Kr$ is the same and it is $k_1$.  Then we must be careful so as not to create a cycle colored $k_1$ within $G_2$. We can notice that
if $e_3$ is colored $k_1$, then within $G_1$ $t_3$ and $t_1$ will also have to be colored with $k_1$ and if we color $e_4$ with $k_1$, then we will have to color $t_4$ and $t_5$ with $k_1$ (and then  $t_3$ and $t_1$  are both colored with two colors of $\Kr'$). Still, one of these options does not create a cycle colored $k_1$ as it cannot be the case that $G_1$ contains two paths colored $k_1$: one beginning at $v_6$ and ending at $v_4$ and the other beginning at $v_6$ and ending at $v_5$.

{\bf Case 2:   Both $e_1$ and $e_2$  are colored with $b'$.}\\
W.l.o.g. we may assume that $b'=b$, $e_3$ is colored $c,d$ and $e_2$  is not colored with $d$. We color $t_4$ with $b$, $t_5$ with $d$ and $t_2$ and $t_3$ with $a$. If $e_2$ is colored with $a$, then we remove $a$ from its coloring and replace it with a color of $\Kr$.
We color $e_4$ with two colors of $\Kr' \setminus \{a,a'\}$. If both $e_2$ and $e_4$ are colored with $c$, then we remove it from the coloring of $e_2$ and replace with a color of $\Kr'$.
Additionally we will either color $t_2$ with $c$ or $t_3$ with $d$.

If $e_3$  (or $e_4$) has two edges coincident with it  at $v_5$ (corr. at $v_4$), then we proceed analogously and replace one of the colors from the coloring of $e_3$ with an appropriate coloring of $\Kr$.
It may happen that it is not possible, because both $G_1$ and $G_2$ contains four edges going out of $v_5$ but then each color with which it is colored is safe, moreover there is a flexibility in chhosing the pair of volors with which $e_3$ can be colored.  

\koniec

{\noindent \bf Acknowledgements} I would like to thank Khaled Elbassioni and Anke van Zuylen for many helpful discussions and Bartek Rybicki for reading the preliminary version of this paper.

\end{document}